\documentclass[10pt]{article}

\usepackage{lipsum}
\usepackage{graphicx}
\usepackage{xcolor}
\usepackage{amsmath} 
\usepackage[square,numbers,sort&compress]{natbib}   
\usepackage[colorlinks=true,linkcolor=blue,citecolor=blue,urlcolor=blue]{hyperref}
\usepackage[french,english]{babel}
\usepackage[T1]{fontenc}
\usepackage[utf8]{inputenc}
\usepackage{minitoc}
\makeatother
\usepackage{enumitem}
\usepackage{mathtools}
\usepackage{amssymb}
\usepackage{enumitem}
\usepackage{adjustbox}
\usepackage{microtype}
\usepackage[labelfont=bf]{caption}
\usepackage{subcaption}
\usepackage{fix-cm}
\usepackage[a4paper, width=140mm, top=25mm,bottom=25mm,bindingoffset=2mm]{geometry}
\usepackage{array}
\usepackage{xfrac}
\usepackage{hypcap}
\usepackage{booktabs}
\usepackage[utf8]{inputenc}  % Set the input encoding to UTF-8
\usepackage{graphicx}
\usepackage{placeins}

 \usepackage{libertine,libertinust1math} 
\usepackage{titlesec}
\titleformat{\section}[block]{\normalfont\large\bfseries}{\thesection}{1em}{}
\date{}
\title{\textbf{\Large{ Measurement, self-similarity, and TNT equivalence of blasts from exploding wires }}}

\author{
  Ahmad Morsel\textsuperscript{1}, 
  Filippo Masi\textsuperscript{2}, 
  Panagiotis Kotronis\textsuperscript{1},
  Ioannis Stefanou\textsuperscript{1}
}
\date{}
\usepackage[pagewise]{lineno}
\begin{document}
\maketitle

\vspace{-10pt}
\begin{center}
\vspace{-10pt}
\small
    $^{1}$Nantes Université, Ecole Centrale Nantes, CNRS,\\
    Institut de Recherche en Génie Civil et Mécanique (GeM), UMR 6183,\\
    Nantes, France. \vspace{3pt}\\
$^{2}$Sydney Centre in Geomechanics and Mining Materials (SciGEM),\\
School of Civil Engineering, The University of Sydney,\\
Sydney, Australia. 
\end{center}
\vspace{10pt}

\begin{abstract}

\noindent Reduced-scale experiments offer a controlled and safe environment for studying the effects of blasts on structures. Traditionally, these experiments rely on the detonation of solid or gaseous explosive mixtures, with only limited understanding of alternative explosive sources. This paper presents a detailed investigation of the blasts produced by exploding aluminum wires for generating shock waves of controlled energy levels. We meticulously design the experiments to ensure a precise quantification of the underlying uncertainties and conduct comprehensive parametric studies. 

We draw practical relationships of the blast intensity with respect to the stand-off distance and the stored energy levels. The analysis demonstrates self-similarity of blasts with respect to the conventional concept of the scaled distance, a desirable degree of sphericity of the generated shock waves, and high repeatability. Finally, we quantify the equivalence of the reduced-scale blasts from exploding wires with high explosives, including TNT. The present experimental setup and study demonstrate the high degree of robustness and effectiveness of exploding aluminum wires as a tool for controlled blast generation and reduced-scale structural testing.\\


\end{abstract}

%% main text
\section{Introduction}
The investigation of the dynamic response of structures subjected to blast loading is a major issue in protecting such assets against explosions. Experimental testing plays a crucial role in enhancing our understanding of how structures respond to blast waves, and it can be broadly classified into full-scale and reduced-scale testing (see \citep{zyskowski2004study,pereira2015masonry,keys2017experimental, Fouchier2017,sochet2018blast,Gebbeken2018,godio2021experimental,morsel2024miniblast}, among others). However, full-scale testing is non-trivial due to the inherent costs, difficulties in ensuring repeatability, and safety risks \citep{draganic2018overview}. In contrast, reduced-scale testing provides increased control and repeatability, with reduced hazards and costs, hence it offers the possibility of performing parametric studies, rather than single tests \citep{farrimond2023microblast,morsel2024miniblast}.\\

In reduced-scale experiments, two critical aspects must be considered: (1) scaling laws and (2) the type of the explosive source. Scaling laws ensure the equivalence between blast loads acting on the reduced-scale model and those acting on the full-scale prototype. Depending on the phenomena under investigation, different scaling laws for blast waves, blast loads, and their effects on structures are available in the literature (see \citep{hopkinson1915british, cranz1925lehrbuch,kinney2013explosive,wei2021new,masi2020scaling}, among others). Regardless of the chosen scaling law, reduced-scale tests typically involve significantly smaller amounts of energy compared to full-scale, field tests. The required explosive mass depends on the type of explosive, its chemical composition, and the dimensions of the tested environment and model, which are governed by the chosen scaling laws. Explosives can vary widely as far as it concerns their chemical composition, e.g., trinitrotoluene (TNT), pentaerythritol tetranitrate (PETN), Composition C4, and are classified as solid or fluid (cf. \citep{meyer2016explosives, zapata2021chemical, needham2018, sochet2018blast}). In reduced-scale settings, solid explosives, including TNT, are often disregarded due to (1) the difficulty in handling and preparing very small explosive masses (as small as a few micrograms, cf. \citep{nof2017exploration}), (2) the need for an igniter to detonate the explosive, and (3) the safety concerns associated with storage which render difficult their use in a laboratory setting.

An alternative consists of gaseous explosives which can be stored more safely, allow precise control of the chemical components for blasts of varying intensities, and are easier to handle in a reduced-scale environment. A common approach consists of detonating gas bubbles initiated electrically -- for a detailed review, we refer to \citep{zyskowski2004study,sochet2018blast}. However, gaseous explosive mixtures may require relatively large gas volumes, posing practical challenges when conducting reduced-scale experiments on centimeter-scale structures while maintaining geometric similarity in terms of stand-off distance and explosive radius.

Another experimental approach consists of resorting to shock tubes for generating shocks in air for tests at medium or small-scale testing \citep{keys2017experimental,schneider2020characterization}. However, the setup produces shock waves that differ from those generated by conventional explosions (with longer characteristic times) and is limited to the study of isolated structural elements as the loading acts along a distinct direction rather than (hemi-)spherically.\\

Alternatively, the use of analogue explosives has been gaining increased attention in reduced-scale testing. This is the case of exploding wires, where a rapid discharge of high electrical loads over a thin conductor produces repeatable blast-type shock waves of controlled intensity similar to those produced by conventional explosives (cf. \citep{chace1964exploding, faraday1857division, Liverts_wires, liu2019experimental, Ram_wires, han2020experiments,  gilburd2012modified}). The concept of exploding wire dates back to the late 18\textsuperscript{th} century, with the seminal work of \citet{nairne1774electrical} who connected slender pieces of silver and copper to a Leyden jar, causing the wire to explode upon discharge. \citet{faraday1857division} later built on this research, using exploding wires to deposit a thin gold film and provided early insights into the phenomenon. Recently, the study of the generation of shock waves from exploding wires gathered further attention (see \citep{bennett1958cylindrical,doney2010experiments,liu2019experimental, mellor2020design, sochet2018blast,Sadot2018}).

Several experimental works have been focused on the understanding of the effects of different parameters affecting the blast generated from exploding wires. For instance, \citet{bennett1958cylindrical} investigated the shape of shock waves generated by exploding wires, using mirrors and high speed cameras, finding cylinder-shaped shock waves fronts. \citet{Ram_wires} investigated blast-structure interactions arising from exploding wires. \citet{Liverts_wires} implemented the exploding wire technique to study blast-wave interaction in aqueous foams. \citet{sochet2018blast} performed an extensive study on the impact of the wire's diameter, length, and capacitor charge on the resulting shock-wave pressure profiles and time-history. In particular, they demonstrated a significant influence of the wire diameter on the resulting blast overpressure. \citet{liu2019experimental} showed that the energy deposition within conductive wires strongly depends on the (dis)charging voltage and the wire dimensions. More recently, \citet{mellor2020design} studied the temporal evolution of shock waves, by means of the Schlieren flow visualization technique, and identified a spatial transition from ellipsoidal to spherical shock wave fronts as they expand outward. \citet{han2020experiments} further investigated the dynamic formation of plasma in copper/nickel alloy exploding wires, showing that the material properties impact the plasma formation.

Despite the aforementioned seminal works and studies, our understanding of the blast loads generated from exploding wires remains limited compared to conventional explosives like TNT. This is particularly true in terms of how these blasts scale with key parameters such as stand-off distance and the energy discharged through the electrical circuit. While detailed characterizations are available for TNT, data on exploding wires remains fragmented and sparse, ultimately hindering their application in studying blast effects on structures within controlled laboratory conditions.\\

The objective of this paper is to provide a comprehensive characterization of the blasts generated from exploding aluminum wires using a newly developed reduced-scale experimental setup for structures, referred to as miniBLAST \citep{morsel2024miniblast,morsel2024fast}. The setup allows us to analyze the effect of varying energies stored in the capacitor before the electrical discharge and the spatial and temporal evolution of the resulting blasts. Therefore, based on experimental measurements of blast overpressure time histories, we compute the corresponding blast impulse, which is crucial for analyzing the effects of an explosion on structures under appropriate scaling laws (cf. \citep{masi2020scaling}). The high repeatability of the experiments provides a robust characterization of the blast parameters, including the arrival time of the shock wave, the positive and negative overpressure peaks, the (positive and negative) load duration, as well as the impulse peaks. Based on this analysis, we demonstrate that the blasts produced from exploding wires follow the same scaling law observed for conventional high-explosives, e.g., TNT \citep{hopkinson1915british,cranz1925lehrbuch}. Next, we compare the blasts generated from exploding wires with conventional solid explosives and draw equivalence factors with TNT charges (cf. \citep{esparza1986blast}). Finally, we illustrate how the generated reduced-scale blasts can be up-scaled to simulate large-scale explosions with the ultimate objective of studying the response and failure of structures subjected to explosions.\\

The paper is structured as follows. Section \ref{sec:methods} outlines the experimental setup and the metrology used to characterize the generated blasts. Particular emphasis is given to the repeatability, sampling rate and measurement errors. Section \ref{sec:results} presents the profile and the time-history of the measured incident overpressure and impulse, at varying of the stand-off distance, $D$, and stored energy, $E_{\textrm{c}} $, in the capacitor. This is followed by a study of the shape of the front of the generated shock waves. Next, Section \ref{sec:blast_params} presents the evolution of the blast parameters as function of the stand-off distance and stored energy, by further discussing the scaling of such parameters with respect to the conventionally-adopted scaled distance and provides best-fit analytical interpolations (regression). Finally, Section \ref{sec:TNT_Equivalence} discusses and presents the TNT equivalence of the generated blasts and offers new insights on the upscaling of reduced-scale explosions from exploding wires.

\section{Methods}
\label{sec:methods}
\begin{figure}[ht]
\centering
\includegraphics[width=\textwidth]{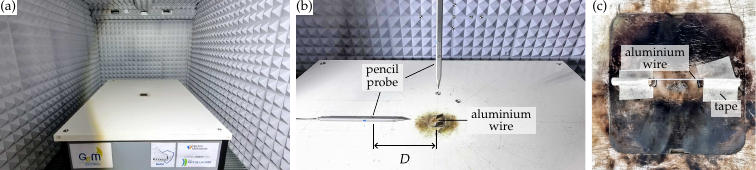}
        \caption{miniBLAST: (a) view from inside the container cabin, (b) installation of the pencil probes at distance $D$ from the explosive source, and (c) detail of the installation of the aluminum wire.}
        \label{fig:fix_sensor}
\end{figure}

% \subsection{miniBLAST platform}
The present experiments leverage the novel experimental setup miniBLAST which ensures a controlled and isolated environment in a laboratory setting (for more details, we refer to \citep{morsel2024miniblast}). miniBLAST has been developed to study the response of structures subjected to blast loads at reduced scale. The experimental setup ensures safety, repeatability, and precise measurements. It is composed of a container cabin of dimensions $4 \times 2.3 \times 2.2$ m\textsuperscript{3} (length\texttimes width\texttimes height). The cabin isolates the experiments from the rest of the laboratory, providing a controlled environment and protection for personnel and equipment. Inside the cabin, an acoustic foam covering ensures absorption of the generated shock waves, thus reducing the impact of any eventual reflections on the measurements, see Figure \ref{fig:fix_sensor}(a).

The experiments are conducted on an optical table ($120$ mm wide, $180$ mm long, and $30.5$ mm high) with six pneumatic supports to ensure a stable and leveled base for testing. This table is specially designed to withstand the blast loads generated during the explosions \citep{morsel2024miniblast} under negligible deformation. A ventilation system is used to remove metal dust particles produced during the detonation. The exploding wire system comprises two circuits: (1) a charging circuit composed of an electric source, a high-current switch, and twelve capacitors (with nominal capacitance $C=408$ \textmu F), and (2) a discharge circuit comprising six Ignitron switches (Ignitron NL7703 mercury switches) and a coaxial cable ($10.7$ m long) that connects the capacitor to the electrodes. The energy, $E_{\textrm{c}} $, that can be stored within the capacitor is in between $5$ J and $46$ kJ (for more details, see \cite{morsel2024miniblast}). The electrodes are installed within the optical table and connected at 15 mm above the plane with an aluminum wire representing the analogue explosive source. In all considered scenarios, the aluminum wire has a diameter of $0.6$ mm and a length of $3.6$ mm, and is weakly fixed to the electrodes by using a commercial tape, see Figure \ref{fig:fix_sensor}(c).\\

For measuring the incident blast wave resulting from the exploding aluminum wire, we use pencil probes, which enable minimum disturbances of the incident air flow and, therefore, minimize reflections and diffraction of the impinging shock wave. We use 6233AA0050 pencil probes with a full scale output of $\pm 200$ kPa. In most scenarios, we position the pencil probes on the optical table at a stand-off distance $D$ from the aluminum wire and fix them adequately with tape, see Figure \ref{fig:fix_sensor}(b). Additionally, we measure the overpressure along the normal direction of the optical table (vertically) by suspending the pencil probe over the table with a stand. The pressure sensors are connected to a data acquisition system (TraNET FE 404) and to an oscilloscope enabling to report the trigger signal at the moment of closing of the discharge circuit. Finally, to prevent the effect of thermal transients \citep{chan1979thermal} stemming from the exploding wire, which may result in fictitious recording of the overpressure signal (e.g., long-standing negative phase), we employ an aluminum foil applied on top of the sensor. For more details, we refer to \cite{morsel2024miniblast}.\\
It is worth noting that the pencil probes could also be mounted within the optical table to further minimize disturbances in the recorded signals caused by interactions between the primary shock wave and the probe body, e.g., reflections and diffraction \cite{rigby2014blast,rigby2014numerical}. Alternatively, pressure gauges could be employed to measure the reflected overpressure (cf. \cite{farrimond2023microblast}). However, as demonstrated below, the high repeatability of the recorded overpressure signals across varying distances and discharge energies suggests that these disturbances are negligible.

\subsection{Analysis of the signal and post-processing}
\label{subsec:postprocess}
The overpressure signal is recorded at a sampling rate of 5 MS/s (mega-samples per second), which provides high accuracy for capturing the overpressure peak (see \citep{morsel2024miniblast}). The sensor has two main sources of uncertainty: (1) acceleration sensitivity of $\pm 0.20$ kPa/g and (2) a linearity error of $\pm 1.16$ kPa. Thus, the total error of the pencil probe is more than satisfactory for an accurate investigation of the positive and negative phases of the blast as it will be shown below. Additionally, the sensor has a low rise time ($<1$ \textmu s) to reach $90\%$ of its total range ($200$ kPa). The maximum recorded level of overpressure increase is approximately $70$ kPa -- that is, $35\%$ of the full scale output -- within a time interval of approximately $15$ \textmu s. Accordingly, the rise time error is negligible in all measurements.

\begin{figure}[ht]
    \centering
    \includegraphics[width=0.522\textwidth]{ 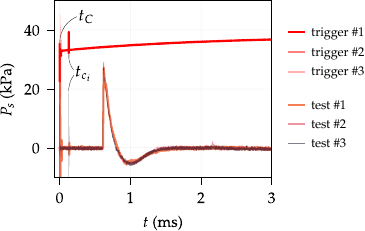}
    \caption{Measurement of the incident overpressure, $P_{\textrm{s}} $, recorded by the pencil probe with sampling rate equal to 5 MS/s, and time evolution of the trigger ($E_{\textrm{c}}  = 5$ kJ and $D = 0.3$ m) from three identical tests. The triggering time, $t_C$, indicates the closure of the switch in the discharge circuit.}
    \label{fig:press_check}
\end{figure}

Figure \ref{fig:press_check} presents the trigger signal and the incident overpressure time-history, shifted to the origin using the triggering time, $t_{\textrm{C}}$, as measured from the oscilloscope. The overpressure signals are recorded from three identical tests at a stand-off distance $D=0.3$ m, for a stored energy $E_{\textrm{c}} =5$ kJ. The signature of the blast load agrees with that resulting from the detonation of solid explosives. Moreover, the three measurements overlap near-perfectly, which demonstrates the excellent repeatability of the generated explosions and the high performance of the measurement system.

It is worth noticing that from the trigger point, $t_{\textrm{C}}$, some minor spikes in the overpressure signal appears at variable times $t_{\textrm{c}_{\textrm{i}}}$, cf. Figure \ref{fig:press_check}. These disturbances are caused by the electrical discharge shortly after the aluminum wire explosion, which results in flashes and occasional electrical arcs. However, in all recorded scenarios, these spikes consistently occur at times less than half of the shock wave arrival time, $t_A$, thus they do not affect the main overpressure measurements.

\begin{figure}[ht]
    \centering
    \includegraphics[width=0.43\textwidth]{ 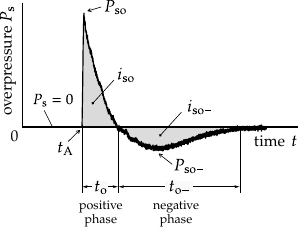}
    \caption{Time evolution of the incident overpressure, $P_{\textrm{s}}$, characterized by the arrival time of the shock wave, $t_{\textrm{A}}$, the peak overpressure, $P_{\textrm{so}} $, the positive phase duration, $t_{\textrm{o}}$, the negative phase duration, $t_{\textrm{o}-}$, and the peak underpressure, $P_{\textrm{so}-} $.}
    \label{fig:pressure_signature}
\end{figure}

For the sake of simplicity, we remove any eventual spikes by post-processing all recorded overpressure measurements by identifying the arrival time, $t_{\textrm{A}}$, and zeroing the value of the pressure for smaller times, $t<t_{\textrm{A}}$. The arrival time is automatically identified as the last intersection point between the recorded signal and the horizontal axis at $P_{\textrm{s}} =0$ kPa, from the trigger time, $t_{\textrm{C}}$, until the point where the overpressure reaches its peak, cf. Figure \ref{fig:press_check}. This results in signals of the type of that schematically sketched in Figure \ref{fig:pressure_signature}. 
With reference to the same figure, we define the conventionally-adopted blast parameters for the positive and the negative phase. For the positive phase, the parameters are: (1) the arrival time, $t_{\textrm{A}}$, (2) the overpressure peak, $P_{\textrm{so}} $, (3) the positive time duration, $t_{\textrm{o}}$, and (4) the positive impulse peak, $i_{\textrm{so}}$. For the negative phase, we define (1) the underpressure peak, $P_{\textrm{so}-} $, (2) the negative impulse peak, $i_{\textrm{so}-}$, and (3) the duration of the negative phase, $t_{\textrm{o}-}$. The peaks $P_{\textrm{so}} $ and $P_{\textrm{so}-} $ are identified as the absolute maximum and minimum values of the pressure signal, respectively. The positive time duration, $t_{\textrm{o}}$, is identified as the intersection between the pressure signal with the horizontal axis $P_{\textrm{s}} =0$ between the peaks $P_{\textrm{so}} $ and $P_{\textrm{so}-} $. Similarly, the negative phase duration, $t_{\textrm{o}-}$, is determined for times $t > t_{\textrm{A}} + t_{\textrm{o}}$. Finally, the impulse, $i_{\textrm{s}}$, and the corresponding peaks, $i_{\textrm{so}}$ and $i_{\textrm{so}-}$, are computed by integrating the overpressure over the relevant time intervals, namely
\begin{equation}
i_{\textrm{s}}(t) =\int_{t_{\textrm{A}}}^{t_{\textrm{A}}+t}P_{\textrm{s}} (t) \, dt, \qquad  
i_{\textrm{so}} = \int_{t_{\textrm{A}}}^{t_{\textrm{A}}+t_{\textrm{o}}} P_{\textrm{s}} (t) \, dt, \qquad
i_{\textrm{so}-} = -\int_{t_{\textrm{A}}+t_{\textrm{o}}}^{t_{\textrm{A}}+t_{\textrm{o}}+t_{\textrm{o}-}}  P_{\textrm{s}} (t) \, dt.
\label{eq:impulse}
\end{equation}
Note that, following the convention in \cite{US_armyTM5}, we define both the negative impulse and underpressure as positive.

\section{Results}
\label{sec:results}
Using the aforementioned setup, we conduct a total of 90 experiments to adequately capture the time history of blasts generated by the exploding wire as a function of the stand-off distance, $D$, and the capacitor's nominal, stored energy, $E_{\textrm{c}} $, with the pencil probe positioned on the optical table, cf. Figure \ref{fig:fix_sensor}, face-on the exploding wire. Next, an additional set of 24 experiments is performed for studying the shape of the front of the generated shock waves.

\subsection{Evolution of the incident overpressure and impulse}
\label{subsec:overpressure_impulse}
The stand-off distance varies between 0.2 and 0.7 m, while the energy levels span between 0.5 and 10 kJ. In parallel, to verify and ensure the repeatability of the experimental results, each test is conducted three times, resulting in three distinct measurements for each scenario.

\begin{figure}[ht!]
\centering
\includegraphics[width=0.83\textwidth]{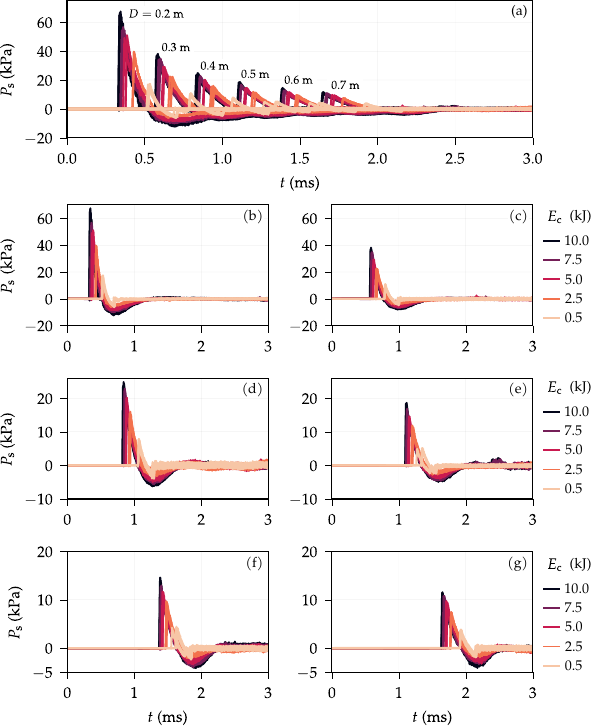}
        \caption{Time evolution of the incident overpressure $P_{\textrm{s}}$ (a) for different energy levels, $E_{\textrm{c}}$, measured at stand-off distances: (b) $D = 0.2$ m, (c) $0.3$ m, (d) $0.4$ m, (e) $0.5$ m, (f) $0.6$ m, (g) and $0.7$ m.}
        \label{fig:pressure_history}
\end{figure}

Figure \ref{fig:pressure_history} presents the time history of the incident pressure for the range of stand-off distances, at varying of the energy level. As expected, higher energy levels $E_{\textrm{c}} $ result in larger amplitudes of the incident overpressure during both the positive and negative phases. Conversely, an increase in the stand-off distance $D$ leads to a reduction in amplitude.

The measurements show lack of parasitic reflections and diffractions, which would affect the time evolution of the measured pressure signal. In all considered scenarios, the positive overpressure is much larger than the error intrinsic to the pencil probe's measurements (approximately $\pm 1$ kPa). However, for very low energies ($E_{\textrm{c}} = 0.5$ kJ) the underpressure peak can sometimes approach the intrinsic error of the sensor. Additionally, we observe a reduced negative phase compared to that at higher energies. This issue is further analyzed in Section \ref{sec:blast_params}.

\begin{figure}[h!]
\centering
\includegraphics[width=0.81\textwidth]{ 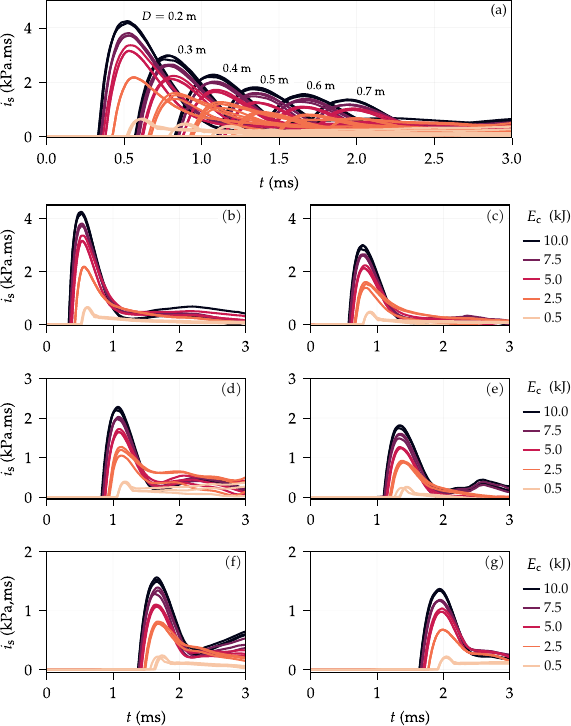}
        \caption{Time evolution of the incident impulse, $i_{\textrm{s}}$, (a) for different energy levels, $E_{\textrm{c}}$, measured at stand-off distances: (b) $D = 0.2$ m, (c) $0.3$ m, (d) $0.4$ m, (e) $0.5$ m, (f) $0.6$ m, (g) and $0.7$ m.}
        \label{fig:impulse_hist}
\end{figure}

Following the definition of the impulse, Equation \ref{eq:impulse}, we compute it from the recorded pressure signals using the trapezoidal rule. The resulting time evolution of the incident impulse is presented in Figure \ref{fig:impulse_hist}. The higher the capacitor's energy $E_{\textrm{c}}$ and the lower the stand-off distance $D$, the larger the impulse is. It is worth mentioning that for low energies, the total impulse -- that is, the value of $i_s$ at the end of the negative phase -- approaches zero. This is due to a non-negligible negative phase, which could be important for the dynamic response of slender structures due to their relatively short characteristic time (cf. \citep{rigby2014negative}).

\subsection{Analysis of the shock wave shape and pressure distribution}
\label{subsec:shock_shape}
\begin{figure}[ht]
    \centering
    \includegraphics[width=0.55\textwidth]{ 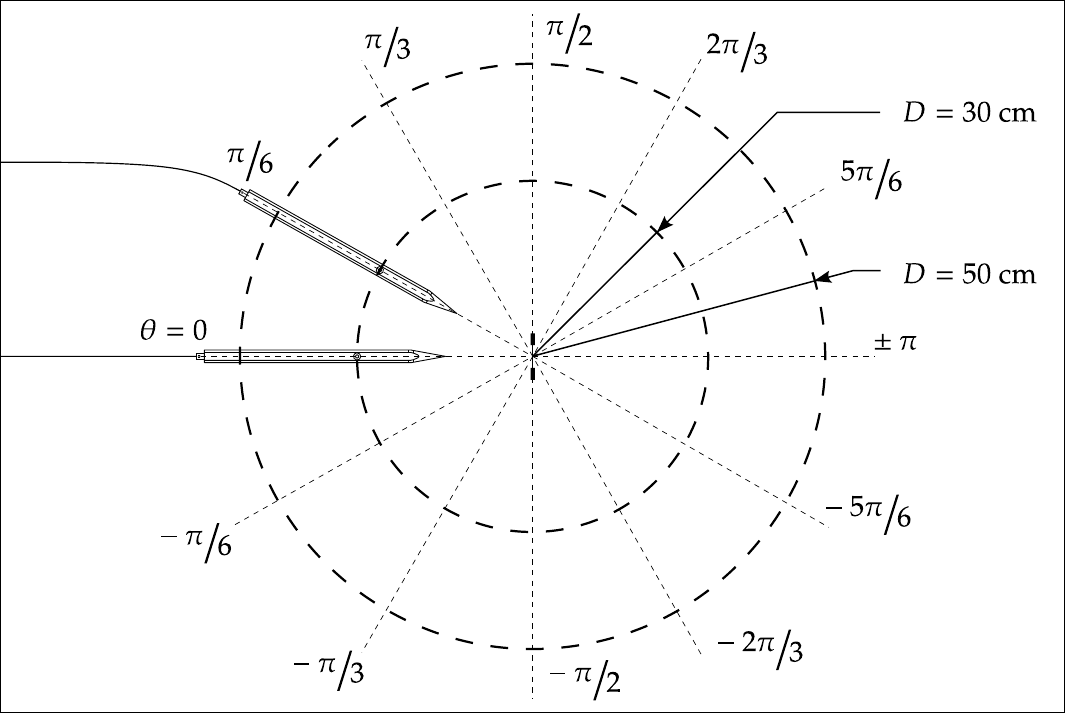}
    \caption{Setup for the study of the shape of the front of the generated shock waves. The pencil probes are positioned at different angles $\theta$ with respect to the normal of the exploding wire, at two stand-off distances, $D$.}
    \label{fig:shock_plane}
\end{figure}

Following the results of \citet{bennett1958cylindrical} and \citet{mellor2020design}, we analyze the sphericity of the shock waves produced by the exploding wire. In particular, we measure the time evolution of the overpressure for different incident angles around the exploding wire position, see Figure \ref{fig:shock_plane}. The pencil probes are positioned at different angles $-\pi\leq\theta\leq \pi$ relative to the direction normal to the axis of the exploding wire.

\begin{figure}[ht]
     \centering
\includegraphics[width=0.88\textwidth]{ 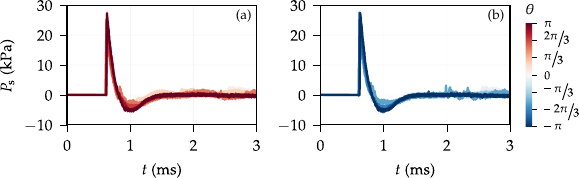}
\caption{Time evolution of the incident overpressure, $P_{\textrm{s}} $, at varying of the incident angle $\theta$ and at a stand-off distance $D=0.3$ m: (a) positive and (b) negative angles, cf. Figure \ref{fig:shock_plane}.}
\label{fig:shape_d30}
\end{figure}

\begin{figure}[ht]
\centering
\includegraphics[width=0.88\textwidth]{ 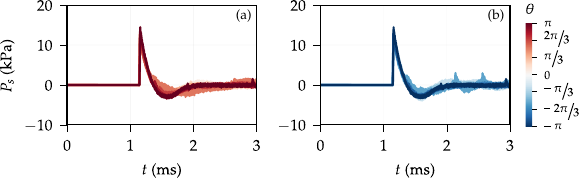}
\caption{Time-history of the incident overpressure, $P_{\textrm{s}} $, at varying of the incident angle $\theta$ and at a stand-off distance $D=0.5$ m: (a) positive and (b) negative angles, cf. Figure \ref{fig:shock_plane}.}
\label{fig:shape_d50}
\end{figure}

The energy level is kept constant, i.e., $E_{\textrm{c}} = 5.0$ kJ, and the tests are performed at two distinct stand-off distances, i.e., 0.3 and 0.5 m. Figures \ref{fig:shape_d30} and \ref{fig:shape_d50} presents the time history of the overpressure with respect to $\theta$, at $D=0.3$ m and $D=0.5$ m, respectively. We observe that the positive phase of the blast is almost identical, independently of $\theta$. Some clearing effects (caused by diffraction) start appearing during the negative phase for angles approaching $\pm \pi /2$ and some reflections are visible at longer times, $t>2$ ms. These reflections and diffraction are likely due to the presence of the electrodes and the shorter length of the optical table.

\begin{figure}[ht!]
    \centering
    \includegraphics[width=0.675\textwidth]{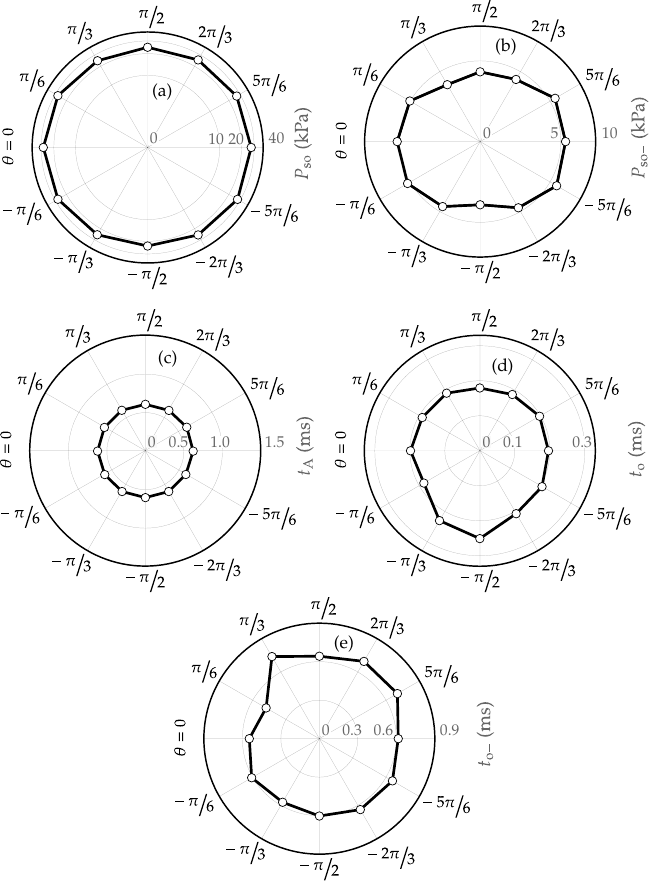}
    \caption{Spatial distribution of the blast parameters at stand-off distance $D=0.3$ m: (a) overpressure peak $P_{\textrm{so}}$, (b) underpressure peak $P_{\textrm{so}-}$, (c) arrival time $t_{\textrm{A}}$, (d) positive phase duration $t_{\textrm{o}}$, and (e) negative phase duration $t_{\textrm{o}-}$, cf. Figure \ref{fig:shock_plane}.}
    \label{fig:shock_shape_03}
\end{figure}

\begin{figure}[ht!]
    \centering
    \includegraphics[width=0.675\textwidth]{ 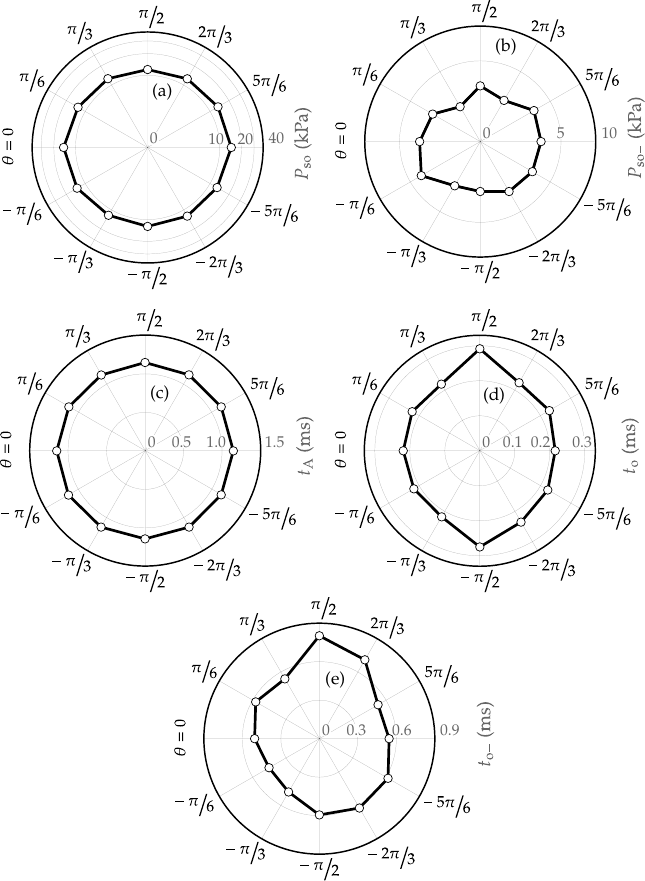}
    \caption{Spatial distribution of the blast parameters at stand-off distance $D=0.5$ m: (a) overpressure peak $P_{\textrm{so}}$, (b) underpressure peak $P_{\textrm{so}-}$, (c) arrival time $t_{\textrm{A}}$, (d) positive phase duration $t_{\textrm{o}}$, and (e) negative phase duration $t_{\textrm{o}-}$, cf. Figure \ref{fig:shock_plane}.}
    \label{fig:shock_shape_05}
\end{figure}

Figures \ref{fig:shock_shape_03} and \ref{fig:shock_shape_05} show the spatial distribution of the (a) overpressure peak, $P_{\textrm{so}} $, (b) underpressure peak, $P_{\textrm{so}-} $, (c) arrival time, $t_{\textrm{A}}$, (d) positive and (e) negative phase duration, $t_{\textrm{o}}$ and $t_{\textrm{o}-}$. Together, these parameters provide a comprehensive approximation of the shock wave front shape.
At very short distances from the wire, the shock wave front is expected to take a cylindrical form (cf. \citep{bennett1958cylindrical}). However, as the distance increases, the front should evolve into a spherical shape (cf. \citep{mellor2020design}). Indeed, in Figures \ref{fig:shock_shape_03} and \ref{fig:shock_shape_05} we observe a nearly spherical (circular) front at $D=0.5$ m and an ellipsoidal front at $D=0.3$ m, with some pronounced boundary effects at $\theta\sim \pi/2$. Note that the spatial distribution of the overpressure peak and arrival time are approximately spherical in both cases. The underpressure peak exhibits an ellipsoidal front with visible boundary effects at $\theta = \pm \pi/2$. Similarly, the positive and negative phase durations display overall circular shock fronts, although some discrepancies arise at angles near $\pm \pi/2$. 
These discrepancies are more pronounced at the larger distance, $D=0.5$ m, and may indicate the presence of reflection and diffraction waves. Although these effects do not impact the primary shock wave (i.e., $P_{\textrm{so}}$ and $t_{\textrm{A}}$), they likely originate from interactions of the primary shock wave with the electrodes and the optical table's edges, which may distort the front's sphericity during the positive and negative phases, i.e., $t_{\textrm{o}}$, $P_{\textrm{so}-}$, and $t_{\textrm{o}-}$.

The sphericity of the primary shock wave front is further confirmed by measurements obtained from a pencil probe suspended above the optical table, see Figure \ref{fig:fix_sensor}(b). In this case, for a stand-off distance $D=0.3$ m, we found that the over- and under-pressure peaks -- $P_{\textrm{so}} =28.1$ kPa and $P_{\textrm{so}-}  = 10$ kPa -- are found to be nearly identical to the values measured at $\theta=0$, at the same distance.

\section{Blast parameters}
\label{sec:blast_params}
\begin{figure}[ht!]
\centering
\includegraphics[width=0.9\textwidth]{ 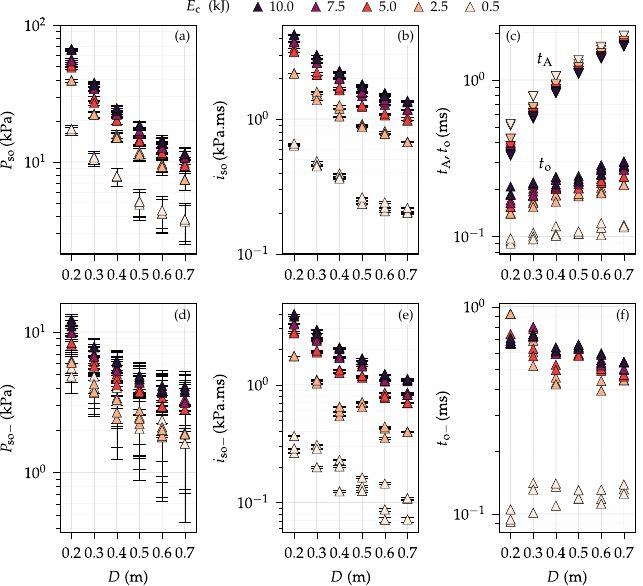}
\caption{Blast parameters versus the stand-off distance, $D$, for different energy levels, $E_{\textrm{c}} $: (a-c) positive and (d-f) negative phases' characteristics. Inset (c) highlights the arrival time, $t_{\textrm{A}}$, with different markers to differentiate it from the positive phase duration, $t_{\textrm{o}}$.}
\label{fig:blast_parameters}
\end{figure}

Following the definitions provided in subsection \ref{subsec:postprocess}, we compute the blast parameters characterizing the positive and negative phases for the case of a normal shock wave, i.e., for $\theta=0$, see Figure \ref{fig:shock_plane}. Figure \ref{fig:blast_parameters} displays how the overpressure peaks, impulses, arrival time, positive and negative phase duration depend on the stand-off distance, $D$, for different energy levels, $E_{\textrm{c}} $. We include error bars associated with the pressure measurements, $P_{\textrm{so}} $ and $P_{\textrm{so}-} $, accounting for the linearity error of the pencil probe, which is $\Delta P = \pm 1.16$ kPa, see subsection \ref{subsec:postprocess}. We neglect the error associated with the acceleration sensitivity of the piezoelectric crystal, which is approximately $\pm 0.20$ kPa/g, because the pencil probe remains fixed throughout all experiments. Indeed, although some accelerations may occur, the corresponding error remains negligible compared to the linearity error (with a difference of one order of magnitude). The error bars for the positive and negative impulses are calculated by propagating the pressure measurement errors, i.e., $\Delta i = \pm \smash{\frac{h\sqrt{n}}{2}}\Delta P$, where $h = 0.2$ \textmu s is the time interval associated with the sampling rate of 5 MS/s, $n$ is the number of samples corresponding to the positive and the negative impulse, $n=t_{\textrm{o}}/h$ and $n=t_{\textrm{o}-}/h$, respectively. The resulting error for $i_{\textrm{so}}$ and $i_{\textrm{so}-}$ is negligible for all scenarios, see Figure \ref{fig:blast_parameters}. No error bars are presented for the arrival time or the duration of the positive and negative phases, as the rise time error of the pencil probe is negligible within the measured pressure range.\\ 
For completeness, Tables \ref{tab1} and \ref{tab2} present the mean value and the standard deviation of the blast parameters, for each scenario, in terms of the energy level, $E_{\textrm{c}}$, and the stand-off distance, $D$.\\

As the energy increases, the overpressure peak, $P_{\textrm{so}} $, the impulse peaks, $i_{\textrm{so}}$ and $i_{\textrm{so}-}$, and time durations, $t_{\textrm{o}}$ and $t_{\textrm{o}-}$, increase. The arrival time presents the opposite trend: the shock wave travels at higher velocity as the stored energy increases, thus $t_{\textrm{A}}$ reduces. For the underpressure peak, we observe that, overall, the higher the energy level is, the higher $P_{\textrm{so}-} $ becomes. However, for low energies, $E_{\textrm{c}} =0.5, 2.5$ kJ, $P_{\textrm{so}-} $ is only slightly affected.

Note that, despite our efforts to accurately capture the negative phase -- such as minimizing thermal transients and secondary reflections -- the measurements for the negative phase are less reliable than those for the positive phase. This is evident from the increasing spread of the error bars at large distances, $D > 0.4$ m, for the underpressure peak, $P_{\textrm{so}-}$. In addition, the negative phase duration, $t_{\textrm{o}-}$, shows two distinct trends depending on the energy level. For $E_{\textrm{c}}>0.5$ kJ, $t_{\textrm{o}-}$ decreases with increasing stand-off distances, regardless of the energy level. In contrast, for $E_{\textrm{c}}=0.5$ kJ, $t_{\textrm{o}-}$ tends to increase with the stand-off distance, with values that are five to six times smaller than those observed at higher energies. This discrepancy may be attributed to two factors. First, at $E_{\textrm{c}}=0.5$ kJ, the intrinsic error of the pencil probe is comparable to the pressure values during the negative phase. Second, the absence of a well-defined negative phase at lower energy levels, as seen in Figure \ref{fig:pressure_history}, could result from an insufficient formation of plasma from the exploding wire to generate a blast-like negative phase. However, the investigation of this phenomenon is beyond the focus of the current work.

\begin{table}[h!]
    \centering
    \caption{Blast parameters (mean value $\pm$ standard deviation) for different energy levels, $E_{\textrm{c}} $, and stand-off distances, $D$: overpressure and underpressure peaks, positive and negative impulse.}
    \label{tab1}
    \small
    \begin{tabular}{ccccccccc}
        \toprule
         % & & \multicolumn{4}{c}{Blast Parameters (mean $\pm$ std)} \\        
        $E_{\textrm{c}}$ & $D$ & $P_{\textrm{so}}$ & $P_{\textrm{so}-}$ & $i_{\textrm{so}}$ & $i_{\textrm{so}-}$ \\
        (kJ) & (m) & (kPa) & (kPa) & (kPa.ms) & (kPa.ms) \\
        \midrule
        0.5 & 0.2 & $17.2 \pm 0.21$ & $5.55 \pm 0.58$ & $ 0.64\pm 0.009$ & $0.31 \pm 0.044$\\
            & 0.3 & $10.7 \pm 0.21$ & $4.21 \pm 0.45$ & $ 0.46\pm 0.017$ & $0.26 \pm 0.047$\\
            & 0.4 & $7.80 \pm 0.04$ & $3.54 \pm 0.34$ & $ 0.38\pm 0.013$ & $0.19 \pm 0.045$\\
            & 0.5 & $5.05 \pm 0.13$ & $2.34 \pm 0.20$ & $ 0.25\pm 0.011$ & $0.14 \pm 0.016$\\
            & 0.6 & $4.26 \pm 0.12$ & $1.98 \pm 0.27$ & $ 0.22\pm 0.013$ & $0.10 \pm 0.032$\\
            & 0.7 & $3.75 \pm 0.06$ & $1.80 \pm 0.14$ & $ 0.21\pm 0.007$ & $0.09 \pm 0.017$\\
        \midrule
        2.5 & 0.2 & $39.5 \pm 0.01$ & $6.07 \pm 0.21$ & $ 2.17\pm 0.023$ & $ 1.75\pm 0.02$\\
            & 0.3 & $22.3 \pm 0.17$ & $3.91 \pm 0.25$ & $ 1.50\pm 0.087$ & $1.08 \pm 0.035$\\
            & 0.4 & $15.5 \pm 0.40$ & $2.80 \pm 0.37$ & $ 1.17\pm 0.091$ & $0.59 \pm 0.044$\\
            & 0.5 & $11.6 \pm 0.22$ & $2.46 \pm 0.22$ & $ 0.89\pm 0.018$ & $0.69 \pm 0.029$\\
            & 0.6 & $9.48 \pm 0.21$ & $1.98 \pm 0.11$ & $ 0.79\pm 0.017$ & $0.41 \pm 0.038$\\
            & 0.7 & $7.41 \pm 0.04$ & $1.87 \pm 0.13$ & $ 0.68\pm0.010 $ & $0.39 \pm 0.022$\\
        \midrule
        5.0 & 0.2 & $50.3 \pm 0.83$ & $ 8.28\pm 0.12$ & $ 3.22\pm 0.093$ & $ 2.78\pm 0.047$\\
            & 0.3 & $28.5 \pm 0.98$ & $ 6.02\pm 0.23$ & $ 2.16\pm 0.051$ & $ 1.92\pm 0.031$\\
            & 0.4 & $20.1 \pm 0.08$ & $ 4.61\pm 0.33$ & $ 1.67\pm 0.035$ & $ 1.31\pm 0.035$\\
            & 0.5 & $14.4 \pm 0.17$ & $ 3.82\pm 0.11$ & $ 1.25\pm 0.013$ & $ 1.18\pm 0.021$\\
            & 0.6 & $11.5 \pm 0.09$ & $ 3.11\pm 0.20$ & $ 1.08\pm 0.015$ & $ 0.81\pm 0.035$\\
            & 0.7 & $9.49 \pm 0.17$ & $ 3.08\pm 0.19$ & $ 0.99\pm 0.027$ & $ 0.70\pm 0.002$\\
        \midrule
        7.5 & 0.2 & $55.3 \pm 0.95$ & $9.91 \pm 0.09$ & $3.74 \pm 0.035$ & $3.30 \pm 0.034$  \\
            & 0.3 & $34.5 \pm 0.22$ & $ 6.79\pm 0.13$ & $ 2.61\pm 0.016$ & $2.45 \pm 0.060$ \\
            & 0.4 & $22.5 \pm 0.28$ & $ 5.46\pm 0.15$ & $ 1.99\pm 0.021$ & $1.73 \pm 0.025$\\
            & 0.5 & $16.5 \pm 0.33$ & $ 4.47\pm 0.12$ & $ 1.55\pm 0.049$ & $1.46 \pm 0.026$\\
            & 0.6 & $12.8 \pm 0.13$ & $ 3.82\pm 0.09$ & $ 1.34\pm 0.046$ & $1.09 \pm 0.019$ \\
            & 0.7 & $10.8 \pm 0.02$ & $ 3.20\pm 0.10$ & $ 1.17\pm 0.008$ & $0.83 \pm 0.018$  \\
        \midrule
        10 & 0.2 & $66.7 \pm 0.59$ & $ 11.7\pm 0.54$ & $ 4.21\pm 0.028$ & $ 3.92\pm 0.078$  \\
            & 0.3 & $37.2 \pm 0.66$ & $ 7.81\pm 0.03$ & $ 2.92\pm 0.083$ & $ 2.85\pm 0.090$ \\
            & 0.4 & $24.2 \pm 0.50$ & $ 6.15\pm 0.07$ & $ 2.24\pm 0.036$ & $ 2.05\pm 0.029$ \\
            & 0.5 & $18.5 \pm 0.13$ & $ 4.77\pm 0.13$ & $ 1.78\pm 0.030$ & $ 1.64\pm 0.031$ \\
            & 0.6 & $14.2 \pm 0.31$ & $ 3.93\pm 0.17$ & $ 1.52\pm 0.032$ & $ 1.22\pm 0.011$ \\
            & 0.7 & $11.5 \pm 0.08$ & $ 3.91\pm 0.15$ & $ 1.35\pm 0.012$ & $ 1.1\pm 0.022$  \\
        \bottomrule
    \end{tabular}
\end{table}

\begin{table}[h!]
    \centering
    \caption{Blast parameters (mean value $\pm$ standard deviation) for different energy levels, $E_{\textrm{c}} $, and stand-off distances, $D$: arrival time, positive and negative phase durations.}
    \label{tab2}
    \small
    \begin{tabular}{ccccc}
        \toprule
         % & & \multicolumn{3}{c}{Blast Parameters (mean $\pm$ std)} \\        
        $E_{\textrm{c}}$ & $D$ & $t_{\textrm{A}}$ &  $t_{\textrm{o}}$ &  $t_{\textrm{o}-}$ \\
        (kJ) & (m) & (ms) &  (ms) &  (ms) \\
        \midrule
        0.5 & 0.2 & $0.518 \pm 0.005$& $0.093 \pm 0.003$ & $ 0.097\pm 0.006$\\
            & 0.3 & $0.785 \pm 0.002$ & $0.100 \pm 0.004$ & $ 0.125\pm 0.017$ \\
            & 0.4 & $1.057 \pm 0.001$ & $0.107 \pm 0.007$ & $ 0.128\pm 0.013$ \\
            & 0.5 & $1.313 \pm 0.037$ & $0.105 \pm 0.002$ & $ 0.122\pm 0.005$ \\
            & 0.6 & $1.599 \pm 0.034$ & $0.113 \pm 0.008$ & $ 0.120\pm 0.008$ \\
            & 0.7 & $ 1.912\pm0.003 $ & $ 0.116\pm 0.002$ & $ 0.132\pm 0.005$ \\
        \midrule
        2.5 & 0.2 & $0.421 \pm 0.001$ & $ 0.141\pm 0.007$ & $ 0.924\pm 0.052$ \\
            & 0.3 & $ 0.662\pm 0.007$ & $ 0.162\pm 0.006$ & $ 0.603\pm 0.068$ \\
            & 0.4 & $ 0.929\pm 0.008$ & $ 0.178\pm 0.008$ & $ 0.436\pm 0.014$ \\
            & 0.5 & $1.193 \pm 0.008$ & $ 0.185\pm 0.003$ & $ 0.599\pm 0.021$ \\
            & 0.6 & $1.473 \pm 0.007$ & $ 0.196\pm 0.006$ & $ 0.445\pm 0.045$ \\
            & 0.7 & $1.768 \pm 0.003$ & $ 0.214\pm0.001 $ & $ 0.442\pm 0.054$ \\
        \midrule
        5.0 & 0.2 & $0.372 \pm 0.007$ & $ 0.159\pm 0.007$ & $0.691 \pm 0.032$ \\
            & 0.3 & $0.614 \pm 0.003$ & $ 0.189\pm 0.009$ & $ 0.663\pm 0.088$ \\
            & 0.4 & $0.876 \pm 0.001$ & $ 0.204\pm 0.002$ & $ 0.541\pm 0.045$ \\
            & 0.5 & $1.143 \pm 0.003$ & $ 0.203\pm 0.000$ & $ 0.601\pm 0.023$ \\
            & 0.6 & $1.418 \pm 0.005$ & $ 0.226\pm 0.005$ & $ 0.503\pm 0.010$ \\
            & 0.7 & $1.706 \pm 0.007$ & $ 0.253\pm 0.017$ & $ 0.467\pm 0.004$ \\
        \midrule
        7.5 & 0.2 & $ 0.346\pm 0.003$ & $ 0.178\pm 0.010$ & $0.669 \pm 0.011$ \\
            & 0.3 & $ 0.586\pm 0.006$ & $ 0.197\pm 0.012$ & $0.750 \pm 0.037$ \\
            & 0.4 & $ 0.846\pm 0.002$ & $ 0.223\pm 0.009$ & $ 0.602\pm 0.007$ \\
            & 0.5 & $ 1.110\pm 0.002$ & $ 0.222\pm 0.005$ & $ 0.632\pm 0.006$ \\
            & 0.6 & $ 1.390\pm 0.004$ & $ 0.266\pm 0.011$ & $ 0.536\pm 0.014$ \\
            & 0.7 & $ 1.662\pm 0.003$ & $ 0.273\pm 0.007$ & $ 0.467\pm 0.001$ \\
        \midrule
        10 & 0.2 & $ 0.331 \pm 0.003$ & $ 0.193\pm 0.01$ & $ 0.687\pm 0.015$ \\
            & 0.3 & $ 0.572\pm 0.004$ & $ 0.219\pm 0.008$ & $ 0.728\pm 0.008$ \\
            & 0.4 & $ 0.829\pm 0.005$ & $ 0.231\pm 0.012$ & $ 0.621\pm 0.012$ \\
            & 0.5 & $ 1.100\pm 0.003$ & $ 0.235\pm 0.007$ & $ 0.650\pm 0.009$ \\
            & 0.6 & $ 1.376\pm 0.003$ & $ 0.266\pm 0.015$ & $ 0.574\pm 0.025$ \\
            & 0.7 & $ 1.638\pm 0.002$ & $ 0.296\pm 0.006$ & $ 0.541\pm 0.001$ \\
        \bottomrule
    \end{tabular}
\end{table}

\subsection{Self-similarity}
\label{subsec:scaling}
For high explosives such as TNT and C4, it is common \citep{US_armyTM5} to scale the blast parameters according to the Hopkinson-Cranz \citep{hopkinson1915british,cranz1925lehrbuch} self-similarity law. This law states that blasts from different explosions at the same atmospheric pressure are self-similar if the associated scaled distance $Z$ is the same, where
\begin{equation}
    Z \equiv \frac{D}{\sqrt[3]{E}},
    \label{eq:Ze}
\end{equation}
$D$ is the stand-off distance, also conventionally referred to as shock wave radius, and $E$ is the (internal) energy of the explosive mass. 

Here, we aim at demonstrating that the same self-similarity is also valid for blasts generated by exploding wires. In doing so, we consider that the internal energy of the ``equivalent'' explosive mass coincides with the energy stored in the capacitor, thus assuming that the whole stored energy is discharged and transferred to the blast, $E\equiv E_{\textrm{c}} $. Note that this is, in general, not necessarily true, as a small part of the stored energy may be consumed or be trapped within the discharge circuit. Moreover,  part of the discharged energy is lost during the phase transition process of the aluminum wire and in the formation of plasma \citep{morsel2024miniblast}. However, considering $E\equiv E_{\textrm{c}} $ allows for a straightforward characterization of the blast parameters with respect to the scaled distance.

\begin{figure}[ht]
\centering
\includegraphics[width=0.9\textwidth]{ 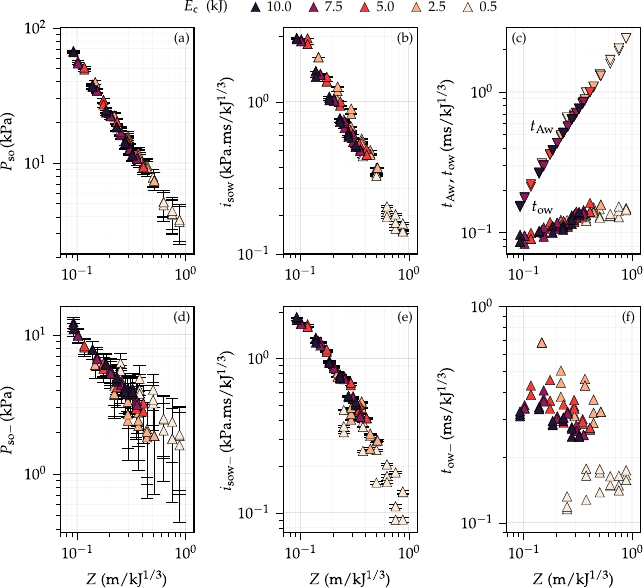}
    \caption{Self-similarity of the blast parameters versus the scaled distance $Z$: (a-c) positive and (d-f) negative phase's parameters.}
    \label{fig:blw_ze}
\end{figure}

Figure \ref{fig:blw_ze} presents the blast parameters as function of the scaled distance, $Z$. The impulse peaks and time durations have been converted to their respective scaled values: scaled impulse peaks, $i_{\textrm{sow}}$ and $i_{\textrm{sow}-}$, and scaled times, $t_{\textrm{Aw}}$, $t_{\textrm{ow}}$, and $t_{\textrm{ow}-}$. The scaling is achieved by dividing each parameter by $\sqrt[3]{E}$, e.g., $i_{\textrm{sow}} = i_{\textrm{so}}/\sqrt[3]{E}$.

Despite minor discrepancies at lower energy levels, specifically at $E_{\textrm{c}} =0.5$ kJ, it is clear that nearly all blast parameters scale very well with the scaled distance, $Z$, confirming the validity of the Hopkinson-Cranz self-similarity law for blasts generated by exploding aluminum wires. The only parameter that shows a weaker correlation with $Z$ is the negative phase duration, $t_{\textrm{ow}-}$. This may be due to the inherently lower accuracy in measuring the negative phase, as compared to the positive phase, as well as the difficulty in precisely determining the end of the negative phase at low energy levels. As it follows, we pursue the analysis of the blast parameters avoiding calculations based on the negative phase duration. To further explore its dependency on the scaled distance, a potential direction for future research is to consider higher energy levels, i.e., $E_{\textrm{c}} > 10$ kJ.\\

We believe that the above experimental results are significant not only because they confirm the self-similarity in exploding wire blasts, but also because they enable effective comparisons with other conventional, well-documented explosives such as TNT. This comparison is explored further in Section  \ref{sec:TNT_Equivalence}.

\subsection{Best-fit interpolations}
\label{subsec:bestfit}
\begin{figure}[ht]
\centering
\includegraphics[width=0.9\textwidth]{ 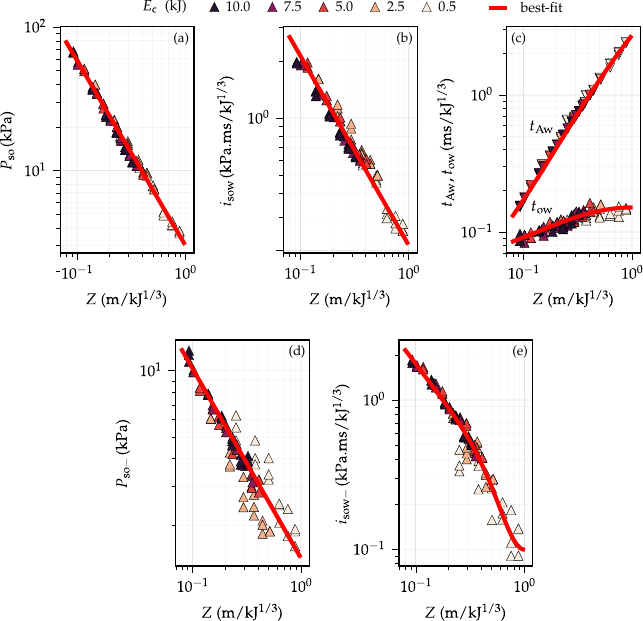}
    \caption{Best-fit interpolations and raw data for (a,d) overpressure, $P_{\textrm{so}} $ and $P_{\textrm{so}-} $, (b,e) scaled impulse, $i_{\textrm{sow}}$ and $i_{\textrm{sow}-}$, and (c,f) scaled arrival time and positive phase duration, $t_{\textrm{Aw}}$ and $t_{\textrm{ow}}$.}
    \label{fig:best_fit}
\end{figure}

We present analytical interpolations of the blast parameters characterizing both the positive and the negative phase. The best-fit interpolations are obtained through symbolic regression, leveraging the \texttt{gplearn} library \citep{gplearn}. By identifying $x\equiv \log Z$, the expression for the pressure peaks, $P_{\textrm{so}} $ and $P_{\textrm{so}-} $ (in kPa), reads
\begin{subequations}
\begin{align}
P_{\textrm{so}} (x) &= \exp \left( \frac{0.856-x}{0.776}\right),\\
P_{\textrm{so}-} (x) & = \exp \big(0.348-0.858x\big),
\end{align}
\label{eq:int_ps}
\end{subequations}
for the scaled impulses, $i_{\textrm{sow}}$ and $i_{\textrm{sow}-}$ in (kPa.ms/kJ\textsuperscript{1/3}),
\begin{subequations}
\begin{align}
i_{\textrm{sow}}(x) &= \exp\big(-1.545-x\big),\\
i_{\textrm{sow}-}(x) &= \exp \big( \tan\left( -0.162-\cos x\right)\big),
\end{align}
\label{eq:int_isw}
\end{subequations}
and for the scaled times, $t_{\textrm{Aw}}$, and $t_{\textrm{ow}}$ (in ms/kJ\textsuperscript{1/3}),
 \begin{subequations}
\begin{align}
t_{\textrm{Aw}} (x) &= 10^3\exp\big(-6.183+x+0.277\cos x\big),\\
t_{\textrm{ow}} (x) &= \frac{1}{10}\exp\big ( 0.411\cos\left(0.78x \right)\big ),\\
% t_{\textrm{ow}-} (x) &= \exp\big (-\cos\left(\sin x\right)-\sin\left(\cos\left(-0.565x\right)\right)\big ).
\end{align}
\end{subequations}
Figure \ref{fig:best_fit} juxtaposes the (raw) blast parameters and the corresponding best-fit interpolations at varying of the scaled distance, $Z$. %est-fit interpolations for the negative phase duration 

Note that the aforementioned expressions are only valid for the particular range of tested scaled distances, i.e., $0.093\leq Z\leq 0.882$ m/kJ\textsuperscript{1/3}, and for the considered electrical circuit, whose details can be found in \cite{morsel2024miniblast}.

\section{Shock wave radius, TNT equivalence, and upscaling}
\label{sec:TNT_Equivalence}
Several representations of a ``universal'' relationship relating the energy (or mass) of explosives with the resulting blast parameters exist. Two of the most common approaches are (1) the shock wave radius versus arrival time concept ($D-t_{\textrm{A}}$, \citep{hargather2023comparison}) and (2) the TNT equivalence factor \citep{esparza1986blast}.

\begin{figure}[ht]
     \centering
     \includegraphics[width=0.71\textwidth]{ 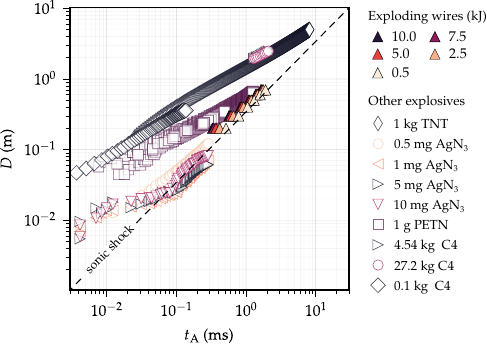}
    \caption{Shock wave radius, $D$, versus arrival time, $t_{\textrm{A}}$, for a range of different explosives and those obtained in this work from exploding wires. The figure is re-adapted from \citep{hargather2023comparison}, where the data related to AgN3 are digitized from \cite{kleine2003studies}, PETN from \cite{winter2021}, C4 from \cite{hargather2013} and \cite{rigby2020}, TNT from empirical formulas \citep{kingery1984,vannucci2017}.}
    \label{fig:D_tA}
\end{figure}

The $D-t_{\textrm{A}}$ relationship allows for the comparison of shock wave profiles generated by different explosives, and facilitates the computation of the shock wave velocity and of the energy. Following \cite{hargather2023comparison}, we compute the $D-t_{\textrm{A}}$ profiles of shock waves driven by exploding wires and compare them with those produced by a range of conventional high explosives, including PETN, TNT, and C4. This comparison is shown in Figure \ref{fig:D_tA}. Note that we show the mean value of the arrival time (over three identical experiments), for each energy level.

The $D-t_{\textrm{A}}$ profiles from the exploding wires, like those of most conventional explosives, lie on the region of supersonic shock waves, indicating that the exploding wire technique enables to obtain shock waves with Mach number $M_{\textrm{s}}>1$, in particular, $1.07 \leq M_{\textrm{s}}\leq 1.8$. The shock profiles produced by exploding wires are comparable to those generated by the detonation of 1 g of PETN, despite a different dependency of the shock velocity on the shock wave radius (slope of the curve).

For completeness, it is worth noting that, following the work of \citet{hargather2023comparison}, the $D-t_{\textrm{A}}$ curves can also be expressed in terms of dimensionless radius and time. Although such a comparison could offer an alternative method for computing the blast energy, this is beyond the scope of the current study and is reserved for future work.

\subsection{TNT equivalence}
\noindent As discussed earlier, different explosives lead to different blast parameters and different dependencies with respect to the scaled distance, $Z$. Under these circumstances, the TNT equivalence is the most widely adopted approach. The goal is to identify the blast parameters arising from the explosive source at hand in terms of TNT, which is historically used as the reference material due to its well-established explosive properties, consistent behavior, and the development of the widely accepted empirical formulae by \citet{kingery1984}.

The TNT equivalence factor $\kappa$ is then defined as the ratio between the energy released by the denotation of a mass $m$ of TNT and the energy associated with the considered explosive, i.e.,
\begin{equation}
    \kappa = \frac{E_{\textrm{TNT}}}{E},
\end{equation}
such that the TNT mass, $m$, and the actual explosive yield the same blast effects, where $E_{\textrm{TNT}}= m \,e_{\textrm{TNT}}$, $e_{\textrm{TNT}} = 4.680$ MJ/kg\textsuperscript{3} is the specific internal energy of TNT, and $E$ is the energy corresponding to the actual explosive.

Several approaches exist to compute the equivalence factor $\kappa$ (for more details, we refer to \citep{sochet2018TNT}). Here, we focus on the most widely adopted and, namely, the equivalence factors based on (1) the positive overpressure peak $P_{\textrm{so}}$, (2) the positive impulse peak $i_{\textrm{so}}$, and (3) the arrival time $t_{\textrm{A}}$ (cf. \citep{esparza1986blast,XIAO2020105871,kleine2003studies,sochet2018TNT}).

According to \cite{esparza1986blast, sochet2018TNT}, we define the three equivalence factors respectively as 
\begin{align}
\kappa_{\textrm{p}} &= \left(\frac{Z}{Z_{\textrm{TNT}}}\right)^3 \quad \text{for} \quad P_{\textrm{so}}  = P_{\textrm{so}} ^{\textrm{TNT}} \label{eq:k_pso}, \\
\kappa_{\textrm{i}} &= \left(\frac{Z}{Z_{\textrm{TNT}}}\right)^3 \quad \text{for} \quad i_{\textrm{so}} = i_{\textrm{so}}^{\textrm{TNT}},\label{eq:k_isow}\\
\kappa_{\textrm{t}} &= \left(\frac{Z}{Z_{\textrm{TNT}}}\right)^3 \quad \text{for} \quad t_{\textrm{A}} = t_{\textrm{A}}^{\textrm{TNT}},\label{eq:k_tA}
\end{align}
where the scaled distance $Z$ is computed according to Equation (\ref{eq:Ze}) and similarly for $Z_{\textrm{TNT}}$, using the energy $E_{\textrm{TNT}}$. 

Following Equations (\ref{eq:k_pso}-\ref{eq:k_tA}), we compute the equivalence factors. For TNT, we leverage the best-fit interpolations of the data collected by \citet{kingery1984}, and for the exploding wires, we consider the raw data points in Figure \ref{fig:blw_ze}.

\begin{figure}[ht]
    \centering
    \includegraphics[width=0.43\textwidth]{ 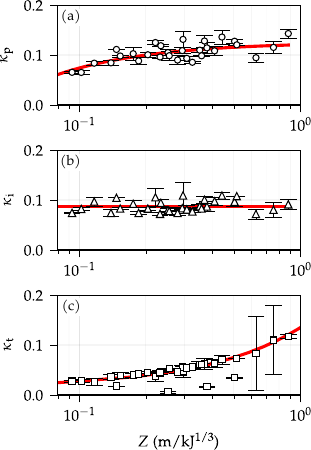}
    \caption{The TNT equivalence factors based on (a) the overpressure peak, $\kappa_{\textrm{p}}$, (b) the scaled positive impulse, $\kappa_{\textrm{i}}$, and (c) the arrival time, $\kappa_{\textrm{t}}$, versus the scaled distance, $Z$. The markers represent the factors obtained using the mean of the raw data, while the error bars display the $95\%$ confidence interval.}
    \label{fig:k}
\end{figure}

Figure \ref{fig:k} displays the three TNT equivalence factors, where the markers represent the mean value obtained from three repeated tests and the bars shows the $95\%$ confidence interval. The three methods -- based on $P_{\textrm{so}} $, $i_{so}$, and $t_A$ -- yield comparable equivalence factors, with corresponding mean values equal to $\kappa_P=0.10$, $\kappa_i=0.088$, $\kappa_t=0.05$. In particular, the overpressure- and impulse-based equivalence factors agree well for $Z\geq 0.12$ m/kJ\textsuperscript{1/3}. The arrival time-based TNT is the smallest for most tested scaled distances and yield similar values with the factors based on the overpressure and the impulse only at large distances, i.e., $Z\geq 0.6$ m/kJ\textsuperscript{1/3} despite the relatively high spread of $\kappa_{\textrm{t}}$. For completeness, we derive best-fit interpolations for each TNT equivalence factor:
\begin{subequations}
\begin{align}
\kappa_{\textrm{p}}(Z) & = \frac{1}{10}\left( 1.26 - \frac{0.52}{10 Z}\right),\\
\kappa_{\textrm{i}} & = 0.088,\\
\kappa_{\textrm{t}}(Z) & = 0.12Z+0.015
\end{align}
\end{subequations}
These expressions are only valid for the particular range of tested scaled distances, i.e., $0.093\leq Z\leq 0.882$ m/kJ\textsuperscript{1/3}, and for the considered electrical circuit.

The equivalence factor based on the impulse, $\kappa_{\textrm{i}}$, is the only factor that remains generally independent of the scaled distance, $Z$. In contrast, the factor based on the overpressure, $\kappa_{\textrm{p}}$, and, in particular, that based on the arrival time, $\kappa_{\textrm{t}}$, exhibit a marked dependence on $Z$. This observation indicates that the commonly-adopted assumption of a single, constant equivalence factor is not appropriate, in general. Similar discrepancies of the equivalence factors, as well as their dependency on $Z$, have been observed in studies of conventional high explosives (see \citep{XIAO2020105871}, among others).

It should also be noted that the values of the equivalence factors may be influenced by the assumed energy associated with the exploding wires. However, even if the effective blast energy, $E$, is lower than the nominal energy stored in the capacitor, i.e., $E < E_{\textrm{c}} $, this would primarily result in a shift toward slightly higher values of $\kappa$. Hence, the difference between $\kappa_{\textrm{p}}$, $\kappa_{\textrm{i}}$, and $\kappa_{\textrm{t}}$, as well as their dependence on scaled distance, would likely remain unchanged.

\subsubsection{TNT equivalent mass}
\begin{figure}[ht]
\centering
    \includegraphics[width=0.9\textwidth]{ 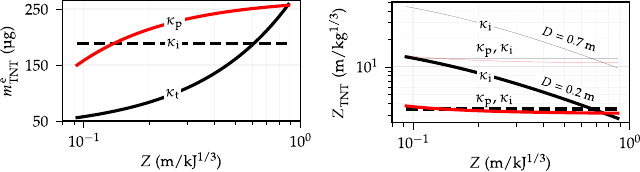}
\caption{TNT equivalent mass, $m^{\textrm{e}}_{\textrm{TNT}}$, and TNT mass-based scaled distance, $Z_{\textrm{TNT}}$ (m/kg\textsuperscript{1/3}), considering a reference energy level $E_{\textrm{c}}= 10$ kJ, obtained using the TNT equivalent factors based on the overpressure peak ($\kappa_{\textrm{p}}$), the impulse peak ($\kappa_{\textrm{i}}$), and the arrival time ($\kappa_{\textrm{t}}$) across varying scaled distances, $Z$.}
        \label{fig:Wpso_isow_e}
\end{figure}

After having determined the equivalence factors, we estimate the equivalent TNT mass, $m^{\textrm{e}}_{\textrm{TNT}}$, whose detonation would produce the same value of the overpressure peak, positive impulse, and arrival at a specified scaled distance, $Z$, according to
\begin{equation}
    m^{\textrm{e}}_{\textrm{TNT}} = \kappa \frac{E_{\textrm{c}} }{e_{\textrm{TNT}} }.
    \label{eq:k_Wtnt}
\end{equation}
Figure \ref{fig:Wpso_isow_e} presents the equivalent TNT mass at varying of the scaled distance, $Z$, for a reference energy level of 10 kJ. For completeness, we also show the relationship between the (energetic) scaled distance, $Z$, and the TNT mass-based scaled distance, $Z_{\textrm{TNT}} = D/m^{\textrm{e}}_{\textrm{TNT}}$.

As expected, the three methods yield different TNT equivalent masses comprised between 50 and 250 \textmu g.

\subsubsection{Comparison of the incident overpressure time histories}
Relying on the determined TNT explosive mass, $m^{\textrm{e}}_{\textrm{TNT}}$, we compare the overpressure time histories recorded during the experiments with those expected from a hemispherical TNT blast. For this comparison, we rely on best-fit interpolations of the data collected by \citet{kingery1984} for the positive phase and negative phase duration. In the absence of specific data and interpolations for the incident underpressure peak, $P_{\textrm{so}-}$, we estimate it by assuming a proportional relationship with the incident overpressure peak, $P_{\textrm{so}}$, analogous to the proportionality between the reflected underpressure and overpressure peaks, which varies with scaled distance $Z$ (cf. \cite{kingery1984}). Accordingly, from the knowledge of the TNT mass and the stand-off distance, we compute the corresponding blast parameters.

\begin{figure}[h]
    \centering
    \includegraphics[width=0.66\textwidth]{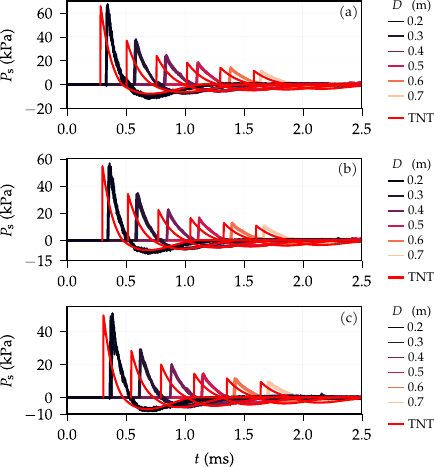}
    \caption{Comparison of the time evolution of the incident overpressure recorded during the experiments with $E_{\textrm{c}}$ equal to (a) 10, (b) 7.5, and (c) 5.0 kJ and that predicted using the overpressure-based TNT equivalence factor, $\kappa_{\textrm{p}}$.}
    \label{fig:comparison_EWp1}
\end{figure}

\begin{figure}[ht]
    \centering
    \includegraphics[width=0.66\textwidth]{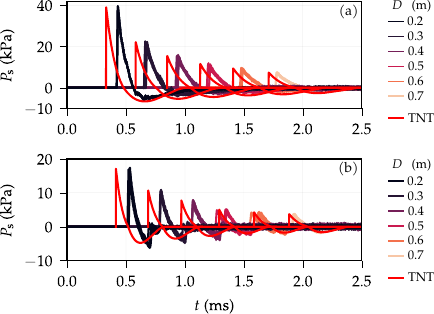}
    \caption{Comparison of the time evolution of the incident overpressure recorded during the experiments with $E_{\textrm{c}}$ equal to (a) 2.5 and (b) 0.5 kJ and that predicted using the overpressure-based TNT equivalence factor, $\kappa_{\textrm{p}}$.}
    \label{fig:comparison_EWp2}
\end{figure}

\begin{figure}[ht]
    \centering
    \includegraphics[width=0.66\textwidth]{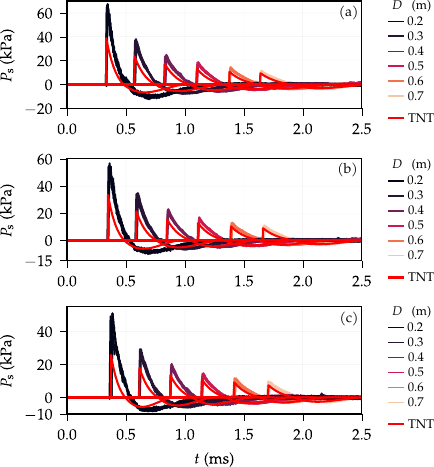}
    \caption{Comparison of the time evolution of the incident overpressure recorded during the experiments with $E_{\textrm{c}}$ equal to (a) 10, (b) 7.5, and (c) 5.0 kJ and that predicted using the arrival time-based TNT equivalence factor, $\kappa_{\textrm{t}}$.}
    \label{fig:comparison_EWt1}
\end{figure}

\begin{figure}[ht]
    \centering
    \includegraphics[width=0.66\textwidth]{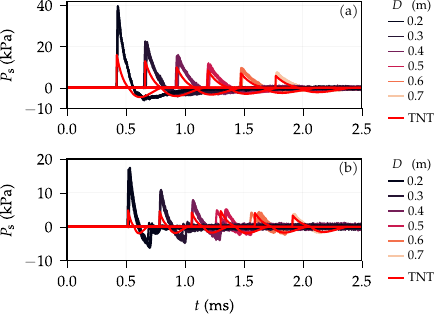}
    \caption{Comparison of the time evolution of the incident overpressure recorded during the experiments with $E_{\textrm{c}}$ equal to (a) 2.5 and (b) 0.5 kJ and that predicted using the arrival time-based TNT equivalence factor, $\kappa_{\textrm{t}}$.}
    \label{fig:comparison_EWt2}
\end{figure}

To model the time evolution of the overpressure, we apply the modified Friedlander equation \cite{friedlander1946diffraction, rigby2014negative} for the positive phase and Granström's cubic equation for the negative phase \cite{cubicnegative,rigby2014negative}.
Figures \ref{fig:comparison_EWp1}-\ref{fig:comparison_EWt2} show the comparison of the resulting overpressure time histories for the overpressure-based and the arrival time-based TNT equivalent mass. In all cases, the comparison confirms the TNT equivalence between the overpressure peak and arrival time, respectively.

From Figures \ref{fig:comparison_EWp1} and \ref{fig:comparison_EWp2}, we observe a time lag in the shock wave arrival when comparing overpressure-based equivalence between exploding wires and ideal TNT blasts. This difference diminishes at larger stand-off distances. Despite the time lag, the overpressure evolution shows good agreement in both the positive and negative phases for energy levels between 5.0 and 10 kJ. In particular, the underpressure peak and negative phase duration closely match. Instead, at lower energy levels ($0.5 \leq E_{\textrm{c}} \leq 2.5$ kJ), some differences appear in the negative phase, especially for 0.5 kJ, where the exploding wire's underpressure exhibits an abrupt discontinuity. This may be due to insufficient level of discharged energy to generate well-defined blast waves.

Figures \ref{fig:comparison_EWt1} and \ref{fig:comparison_EWt2} reveal overall lower overpressure peaks for ideal TNT blasts when using the arrival time-based TNT equivalence. However, at larger stand-off distances ($0.5 \leq D \leq 0.7$ m), the overpressure evolution from TNT and exploding wire blasts is nearly identical across all considered energy levels, with the exception of the $E_{\textrm{c}} = 0.5$ kJ case. This agreement, in contrast with the differences in overpressure and underpressure peaks at shorter distances, suggests that the propagation velocity of blast waves from exploding wires diverges spatially from that of conventional explosives (i.e., TNT), cf. Figure \ref{fig:D_tA}. For $E_{\textrm{c}} = 0.5$ kJ, the comparison remains inconclusive.

\subsection{Blast upscaling}
\label{subsec:upscaling}
It is interesting to explore how the (reduced-scale) blasts generated by exploding aluminum wires can be scaled to represent a full-scale prototype explosion. To achieve this, we leverage the scaling laws developed by \citet{masi2020scaling} for the rigid-body response of structures subjected to blast loads (cf. \cite{masi2019rocking,masi2020discrete}). The scaling laws consider both geometric and mass scaling by means of the geometric scale factor, $\lambda = {D}/{L}$, and the density scale factor, $\gamma = {\tilde{\rho}}/{\rho}$, where $L$ and $\rho$ are the characteristic length and density of the full-scale structure, i.e., the prototype, while $D$ and $\tilde{\rho}$ denote the corresponding reduced-scale structure's quantities, i.e., the model. The scaling is obtained by requiring the self-similarity of the impulse under the assumption of impulsive blast loads. This yields the scaling factor for the TNT mass, $\lambda_m = {m}^{\textrm{e}}_{\textrm{TNT}}/{M}^{\textrm{e}}_{\textrm{TNT}}$, that is obtained by solving the following non-linear equation:
\begin{equation}
  \textrm{find } \lambda_m \textrm{ such  that } \frac{\tilde{i}_{\textrm{ow}}\left(\lambda/\lambda_m^{1/3} Z_{\textrm{TNT}}\right)}{i_{\textrm{ow}}(Z_{\textrm{TNT}})} - \gamma \frac{\lambda^{3/2}}{\lambda_m^{1/3}}=0,
   \label{TNT}
\end{equation}
where $Z_{\textrm{TNT}}$ is the TNT mass-based scaled distance in the prototype, and $\tilde{i}_{\textrm{ow}}$ and ${i}_{\textrm{ow}}$ are respectively the model and the prototype scaled impulse peaks, computed from the best-fit interpolations of TNT blasts. Originally the scaling laws account the reflected impulse; however, here, we consider the incident impulse for the sake of simplicity. Also, notice that when $\lambda_Z=1$, the above equation results in the conventional Hopkinson-Cranz similarity law (for more details, we refer to \cite{masi2020scaling}).\\

\begin{figure}[ht]
    \centering
    \includegraphics[width=0.56\textwidth]{ 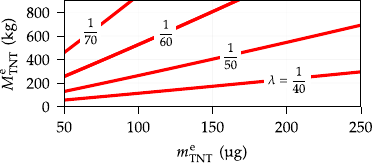}
    \caption{Upscaled TNT equivalent mass $M^{\textrm{e}}_{\textrm{TNT}} $ from the model with mass $m^{\textrm{e}}_{\textrm{TNT}} $, for different values of the geometric scaling, $\lambda$. The TNT equivalence is based on the positive scaled impulse peak.}
    \label{fig:Mp_pso}
\end{figure}

Here, we assume a unit mass scaling and several geometric scaling factors, i.e., $\sfrac{1}{70}\leq \lambda \leq \sfrac{1}{40}$. Figure \ref{fig:Mp_pso} presents the upscaling relationship between the TNT equivalent mass in the prototype, $M^{\textrm{e}}_{\textrm{TNT}}$, and the actual mass in the model, $m^{\textrm{e}}_{\textrm{TNT}}$, for both TNT equivalence factors ($\kappa_{\textrm{p}}$ and $\kappa_{\textrm{t}}$). These results represent the proof of concept for the current study, demonstrating how reduced-scale experiments using exploding wires could be employed to investigate blast effects on structures.

\section{Conclusions}
Reduced-scale experiments offer compelling advantages over full-scale and field experiments, when appropriate scaling laws are available \citep{hopkinson1915british,masi2020scaling}. Existing research has primarily focused on blast experiments using solid or gaseous mixtures of explosive \citep{needham2018, sochet2018blast}. However, these explosive sources pose practical challenges due to environmental hazard and difficulties in managing small explosive charges while ensuring geometric similarity with full-scale prototypes. In response to these limitations, exploding wires have gained attention for their ability to generate controlled, blast-like shock waves \citep{nof2017exploration,Sadot2018,mellor2020design}.

Building on this technology, we conducted an extensive analysis of the blasts generated by exploding aluminum wires using a novel experimental setup designed for studying blast effects on structures in a controlled and safe laboratory environment \citep{morsel2024miniblast}. Through parametric tests and repeated experiments under identical conditions, we demonstrated a high level of robustness and repeatability in the resulting blasts. This analysis provided a comprehensive examination of how blast intensity varies with both the stand-off distance and the energy stored in the capacitor. Next, we identified key blast parameters, including the over- and under-pressure peaks, the positive and negative impulses, as well as the arrival time, and the positive and negative phase duration. Those parameters were found to follow the self-similarity law proposed by \citet{hopkinson1915british} that is commonly applied to high explosives.

Finally, we studied the equivalence of blasts from aluminum wires to those from conventional explosives, such as TNT. We calculated the TNT equivalence factors and established a direct correspondence between the stored energy and the equivalent mass of TNT. These factors were found to be strongly dependent on the particular approach used to evaluate them, namely relying on the overpressure peak or the scaled positive impulse, as well as the stand-off distance. This finding indicates that single, constant equivalence factors may be inappropriate.

The novel experimental setup and the detailed characterization of the blasts from exploding wires open new perspectives in the study of reduced-scale explosions and of the fast, structural dynamics due to blast loads, with reduced costs, increased safety, and repeatability of tests. As a result, this work lays foundations for broader investigations in engineering structures.

\subsubsection*{Declaration of competing interest}
The authors declare that they have no known competing financial interests or personal relationships that could have appeared to influence the work reported in this paper.

\subsubsection*{Data availability}
Experimental data accompanying this manuscript are publicly available at \cite{data}.

\subsubsection*{Acknowledgments}
The authors would like to acknowledge the support of the Region Pays de la Loire and Nantes Métropole under the Connect Talent programme (CEEV: Controlling Extreme EVents - BLAST: Blast LoAds on STructures). Special thanks also go to Prof. Guillaume Racineux and Mr. Emmanuel Marché for our fruitful discussions and their technical assistance regarding the implementation of the system.

\bibliography{sn-bibliography}
\end{document}